\documentclass[showpacs,aps,pra,twocolumn,floatfix,superscriptaddress,nofootinbib]{revtex4}
\usepackage{graphicx}
\usepackage{bm}
\usepackage{amsmath}
\usepackage{euscript}
\usepackage{graphics}
\usepackage{bm}
\usepackage{epsfig}
\usepackage{amssymb}

\newcommand{\be}{\begin{equation}}
\newcommand{\e}{\end{equation}}
\newcommand{\beml}{\begin{subequations}}
\newcommand{\eml}{\end{subequations}}
\newcommand{\beq}{\begin{eqnarray}}
\newcommand{\eq}{\end{eqnarray}}
\newcommand{\ba}{\begin{array}}
\newcommand{\ea}{\end{array}}
\newcommand{\unite}{\mathbf{\hat{e}}}
\newcommand{\lt}{\left}
\newcommand{\rt}{\right}
\newcommand{\n}{\nonumber}
\newcommand{\la}{\langle}
\newcommand{\ra}{\rangle}
\newcommand{\tr}{{\rm Tr}\,}

\newcommand{\re}{\,{\rm Re}\,}
\newcommand{\ep}{\boldsymbol{\varepsilon}}

\newcommand{\Du}{\textbf{D}^{\dagger}}
\newcommand{\Dd}{\textbf{D}}
\newcommand{\tD}{\overleftrightarrow{\boldsymbol{\Delta}}}
\newcommand{\vk}{\textbf{k}}

\newcommand{\bra}[1]{\left|#1\right\rangle}
\newcommand{\ket}[1]{\left\langle#1\right|}

\begin{document}

\date{\today}

\title{Spectrum of coherently backscattered light from two atoms}

\author{\firstname{Vyacheslav} \surname{Shatokhin}}
\affiliation{Max-Planck-Institut f\"ur Physik komplexer Systeme,
N\"othnitzer Str. 38, 01187 Dresden, Germany}
\affiliation{B.~I.~Stepanov Institute of Physics, National Academy
of Sciences, Skaryna Ave. 68, 220072  Minsk, Belarus}
\author{\firstname{Thomas} \surname{Wellens}}
\affiliation{Institut f\"ur Theoretische Physik, Universit\"at
Erlangen-N\"urnberg, Staudstr. 7, 91058 Erlangen, Germany}
\author{\firstname{Beno\^{\i}t} \surname{Gr\'emaud}}
\affiliation{Laboratoire Kastler Brossel, Universit\'e Pierre et
Marie Curie, 4, place Jussieu, 75252 Paris Cedex 05, France}
\author{\firstname{Andreas} \surname{Buchleitner}}
\affiliation{Max-Planck-Institut f\"ur Physik komplexer Systeme,
N\"othnitzer Str. 38, 01187 Dresden, Germany}

\begin{abstract}
We present a detailed analytical and numerical analysis of the
inelastic coherent backscattering spectrum of laser light incident
on cold atoms. We identify frequency domains where the interference
contribution can be positive as well as negative -- or exhibits
dispersive character. These distinctive features are explained by
reciprocity arguments and dressed state two-photon scattering
amplitudes.
\end{abstract}

\maketitle

\section{Introduction}
Multiple scattering of light in cold atomic gases has become an area
of intense theoretical and experimental research (for a recent
review, see \cite{kupriyanov06}). On the experimental side, the
successful observation of coherent backscattering (CBS) of light in
clouds of cold atoms \cite{labeyrie99,kulatunga03} demonstrated the
potential of finely tunable atomic media for detailed studies of
localization and transport phenomena \cite{meso94} in the weak and,
prospectively, strong localization regime. On the theoretical side,
it is of crucial importance to understand how interference effects
are affected by various dephasing mechanisms characteristic for
atom-photon interactions,
 for ensembles of atoms cooled down to approximately 100 $\mu$K, with, in general,
degenerate electronic structure, and in the presence of inelastic
scattering. It is presently also realized that multiple scattering
of light in atomic clouds is relevant in the context of quantum
information storage and retrieval by photons, in the parameter
regime of electromagnetically-induced transparency (EIT)
\cite{lukin03}. This promotes CBS studies toward the dynamical
regime \cite{datsuk06}. Also multiple scattering of {\it
nonclassical} light \cite{skipetrov06} is presently moving into
focus.

In our present contribution, we will expand on the impact of
inelastic scattering processes on CBS. Experimental studies on cold
Sr atoms revealed a rapid decrease of the CBS interference contrast
with increasing intensity of the injected laser field, as a
consequence of the saturation of the laser-driven atomic transition
\cite{chaneliere03}. A much weaker sensitivity of the quality of the
CBS signal was observed for Rb atoms \cite{balik05}. In this latter
case, inelastic processes occur on degenerate transitions
\cite{jonckheere00,mueller01,mueller02,kupriyanov03,labeyrie03},
while Sr atoms offer dipole transitions with a nondegenerate ground
state \cite{bidel02}.

So far, the role of the
 nonlinearity of the atom-photon interaction for CBS scatterers
 has been theoretically investigated only for Sr atoms
\cite{wellens04,shatokhin05,shatokhin06,gremaud06,shatokhin07}.
Within a scattering theoretical approach applied to two atoms in the
regime of weakly nonlinear scattering \cite{wellens04}, the decrease
of the CBS enhancement factor $\alpha$ was shown to be due to the
partial distinguishability of the interfering amplitudes. In the
general case of many atoms, three different amplitudes interfere
constructively in the weakly nonlinear regime, such that $\alpha$
may exceed the linear barrier two \cite{wellens05,wellens06}. For
arbitrary intensities of the injected laser light, a number of
effects have been predicted within a master equation approach
\cite{shatokhin05,shatokhin06}, such as a nonvanishing residual CBS
contrast in the deep saturation regime, or CBS anti-enhancement
under off-resonant driving. The strongly inelastic scattering regime
studied by a quantum Langevin treatment highlighted the crucial role
of inelastic susceptibilities \cite{gremaud06}.

Here, we will focus on the spectral properties of the CBS signal, in
order to provide a detailed interpretation of the residual CBS
enhancement or anti-enhancement predicted earlier
\cite{shatokhin05,shatokhin06}. This will also elucidate the
structure of the CBS spectra presented in
\cite{gremaud06,shatokhin07}. Specifically, we will identify
frequency domains where the interference contribution to CBS
exhibits not only constructive but also destructive terms, or else a
dispersive lineshape. We relate the spectral lines of the CBS
spectrum to CBS transitions between atomic dressed states. We show
that, in the limit of very intense laser fields, spectral ranges
with destructive interference lead to a decreased enhancement
factor, which, at exact resonance, shrinks to $\alpha_\infty=23/21$.
For off-resonance driving, the destructively interfering processes
can outweigh the constructively interfering ones, leading to values
of the enhancement factor less than unity.

The paper is organized as follows: We start with a brief
presentation of our two-atom model, and of the master equation
approach. In Sec.~\ref{sect:results} we present our results on the
CBS spectrum. Section~\ref{sec:conclusion} concludes the paper.

\section{A master equation for two atoms}
\subsection{The model, and the main quantity of interest}
While details of our approach were presented elsewhere
\cite{shatokhin06}, we recollect its basic ingredients relevant for
our subsequent spectral analysis. We start out with the general
formulation of a Hamiltonian describing $N$ identical, stationary
atoms embedded in an electromagnetic environment of quantized
harmonic oscillators, and subjected to an external (classical) laser
field of arbitrary intensity. Coupling to the bath gives rise to
spontaneous emission from the excited state and to the far-field
dipole-dipole interaction responsible for the exchange of photons
between the atoms, whereas the coupling to the laser field induces
Rabi oscillations of the populations and coherences on the
laser-driven atomic transitions.

The intensity of the light scattered off the atomic system is
expressed via the correlation functions of the emitting dipoles. We
address a regime of multiple scattering which is relatively simple
from the theoretical point of view: scattering by an optically thin
atomic medium, where double scattering provides the dominant
contribution to the CBS signal \cite{jonckheere00,bidel02}. It is in
this double scattering regime where the first observation of a CBS
reduction due to the saturation of atomic dipole transitions was
reported \cite{chaneliere03}.

\begin{figure}
\includegraphics[width=8cm]{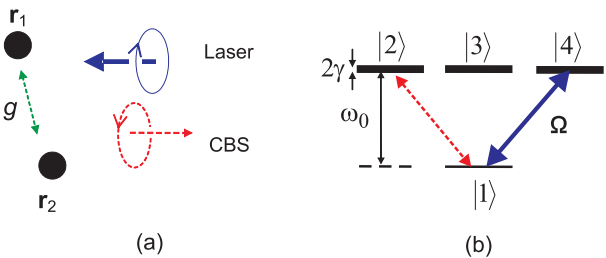}
\caption{ Model of CBS with two atoms. (a) atoms (black dots) are
driven by laser light with right circular polarization, while CBS is
observed in the helicity preserving channel, that is, with flipped
polarization. Photons in this channel appear as a result of double
scattering. $g$ is the strength of the far-field dipole-dipole
coupling responsible for the exchange of photons; (b) internal
atomic structure corresponding to a $J_g=0\rightarrow J_e=1$ dipole
transition. $\omega_0$ is the transition frequency, $2\gamma$ is the
radiative linewidth, $\Omega$ is the Rabi frequency. Sublevels $\bra
1$ and $\bra 3$ have magnetic quantum number $m=0$. Sublevels $\bra
2$ and $\bra 4$ correspond to $m=-1$ and $m=1$, respectively. The
thick solid arrow represents the laser field driving the
$|1\ra\leftrightarrow|4\ra$ transition, while the dashed arrow
indicates the CBS field originating from the
$|1\ra\leftrightarrow|2\ra$ transition. } \label{fig:hph}
\end{figure}

Under this specific conditions, considering a system of only two
atoms suffices to grasp the essential physical phenomena. Thus, we
come up with the toy model of CBS depicted in Fig.~\ref{fig:hph}. We
will study CBS from two identical, motionless atoms, located at
positions ${\bf r}_1$ and ${\bf r}_2$, with the distance
$r_{12}=|{\bf r}_1-{\bf r}_2|$ being much greater than the optical
wavelength $\lambda$.

As for the internal atomic structure we choose nondegenerate atomic
ground states, and the excited state with a three-fold degeneracy
[see Fig.~\ref{fig:hph}(b)], precisely as in the Sr experiment
\cite{chaneliere03}. The laser intensity is encoded in the
saturation parameter $s=\Omega^2/2(\gamma^2+\delta^2)$, where
$\Omega$ is the Rabi frequency, $\gamma$ is half the spontaneous
decay rate of the atomic excited states, and
$\delta=\omega_L-\omega_0$ is the laser-atom detuning.

We will consider the CBS signal in the helicity preserving
($h\parallel h$) polarization channel, as in \cite{chaneliere03},
with right circularly polarized laser light driving the
$|1\ra\leftrightarrow|4\ra$ transition, that is,
$\ep_L=\unite_{+1}$, in helicity basis notation. CBS with preserved
helicity then corresponds to the detection of photons with flipped
polarization,
 $\ep=\unite_{-1}$, that is, from $|1\ra\leftrightarrow|2\ra$ transition,
as shown in Fig.~\ref{fig:hph}.

The CBS spectrum can be derived from the average value of the
first-order temporal correlation function of the field
\cite{glauber65}: \be G^{(1)}({\bf r},t;{\bf
r},t^\prime)=\lt\la\tr\{\rho[\ep\cdot{\bf E}_{\rm s}^{(-)}({\bf
r},t)] [\ep^*\cdot{\bf E}_{\rm s}^{(+)}({\bf
r},t^\prime)]\}\rt\ra_{\rm conf}, \label{corr_func} \e where $\rho$
is the initial density operator of the atom-field system, ${\bf
E}_{\rm s}^{(-/+)}({\bf r},t)$ is the negative/positive frequency
component of the electric field operator of the scattered field, and
$\lt\la\ldots\rt\ra_{\rm conf}$ denotes a configuration average. The
components of the scattered field are the retarded fields radiated
by the atomic dipoles, \be {\bf E}_{\rm s}^{(+)}({\bf r},t) =
\frac{\omega_0^2}{4\pi\varepsilon_0c^2r} \sum_{\alpha=1}^2
\Dd_\alpha(t_\alpha) e^{-i\vk\cdot{\bf r}_\alpha}\, , \label{Eplus}
\e where $\varepsilon_0$ is the permittivity of the vacuum,
$\Dd_\alpha=-\unite_{-1}\sigma^\alpha_{12}+\unite_0
\sigma^\alpha_{13}-\unite_{+1}\sigma^\alpha_{14}$, with
$\sigma^\alpha_{kl}\equiv\bra k_\alpha\ket l_\alpha$, is the dipole
lowering operator, and $t_\alpha=t-|{\bf r}-{\bf r}_\alpha|/c$. In
writing Eq.~(\ref{Eplus}), we have assumed that $r_{12}\ll r$, that
is, the field is detected at a distance much larger than the
interatomic distance. In the following, we will for brevity omit the
$r$-dependent prefactor of Eq.~(\ref{Eplus}) and, consistently, of
the temporal correlation functions.

Inserting Eq.~(\ref{Eplus}) into Eq.~(\ref{corr_func}) we obtain, in
the steady state limit $t\to\infty$, \be G^{(1)}_{\rm
ss}(\tau)=\sum_{\alpha,\beta=1}^2\la\la\sigma^\alpha_{21}\sigma^
\beta_{12}(\tau)\ra_{\rm
ss} e^{i{\bf k}\cdot{\bf r}_{\alpha\beta}}\ra_{\rm conf},
\label{G_tau} \e where ``ss" stands for {\it steady state},
$\tau=t^{\prime}-t\geq 0$, ${\bf r}_{\alpha\beta}\equiv {\bf
r}_\alpha-{\bf r}_\beta$, and the inner angular brackets indicate
the quantum mechanical expectation value [see
Eq.~(\ref{corr_func})].

The spectrum follows via a Laplace transform of (\ref{G_tau})
\cite{scully97}, \be S(\nu)=\frac{1}{\pi}\lim_{\Gamma\to 0}
\re\bigl\{\tilde{G}^{(1)}_{\rm ss}(z)\bigr\}, \label{sp_fixed} \e
where $\tilde{G}^{(1)}_{\rm
ss}(z)=\int_0^{\infty}d\tau\exp(-z\tau)G^{(1)}_{\rm ss}(\tau)$,
 $z=\Gamma-i\nu$, with $\Gamma\geq 0$, and $\nu=\omega-\omega_L$. Note that the
spectrum is defined with respect to the laser frequency, what
implies that the atomic correlation functions must be evaluated in
the frame rotating at $\omega_L$.

\subsection{Configuration average}
A configuration average is necessary because the two-atom
correlation functions may sensitively depend on the interatomic
distance, and on the orientation of the vector ${\bf r}_{12}$ with
respect to ${\bf k}_L$, and thus exhibit rapid oscillations around
the backscattering direction. These oscillations have the same
nature as a speckle pattern scattered off a disordered medium. After
many realizations of the disorder all interference maxima except the
one due to CBS disappear. A simple and sufficient way to mimic
disorder in our two-atom system is to assume an isotropic
distribution of the radius-vector connecting the atoms, and a
uniform distribution (with a width $\lambda$) of interatomic
distances around the average distance $\ell$ equal to the scattering
mean free path.

\subsection{Master equation}
To deduce the atomic correlation functions which enter the right
hand side of Eq.~(\ref{G_tau}) we adapted
\cite{shatokhin05,shatokhin06} the theoretical approach of
\cite{lehmberg70}. Within this setting, the dynamics of the dipole
operators' expectation values as well as of the dipole-dipole
correlators is governed by the master equation \be \la\dot
Q\ra=\sum_{\alpha=1}^2\la{\cal L}_\alpha Q\ra+\sum_{\alpha\neq
\beta=1} ^2\la{\cal L}_{\alpha\beta}Q\ra, \label{meq} \e where the
Liouvillians ${\cal L}_\alpha$ and ${\cal L}_{\alpha\beta}$ generate
the time evolution of an arbitrary atomic operator $Q$, for
independent and interacting atoms, respectively. Explicitly, \beq
{\cal L}_\alpha
Q\!\!\!&=&\!\!\!-i\delta[\Du_\alpha\!\cdot\!\Dd_\alpha,Q]
\!-\!\frac{i}{2}[\Omega_\alpha(\Du_\alpha\!\cdot\!\ep_L)\!+\!\Omega^*_\alpha
(\Dd_\alpha\!\cdot\!\ep_L^*),Q]\nonumber\\
&&+\gamma(\Du_\alpha\!\cdot\![Q,\Dd_\alpha]\!+\![\Du_\alpha,Q]
\!\cdot\!\Dd_\alpha),\label{L_a}
\\
{\cal L}_{\alpha\beta}Q\!\!\!&=&\!\!\!\Du_\alpha\!\cdot\!\overleftrightarrow{\bf
    T}(g,{\bf\hat
n})\!\cdot\![Q,\Dd_\beta]\!+\![\Du_\beta,Q]\!\cdot\!\overleftrightarrow{\bf
T}^*(g,{\bf\hat n})\!\cdot\!\Dd_\alpha\, , \label{L_ab} \eq where
$\Omega_\alpha=\Omega e^{i{\bf k}_L\cdot{\bf r}_\alpha}$ is the
position-dependent Rabi frequency. The radiative dipole-dipole
interaction due to exchange of photons between the atoms is
described by the tensor $ \overleftrightarrow{\bf T}(g,{\bf\hat
n})=\gamma g \overleftrightarrow{\boldsymbol{\Delta}}$, with
$\overleftrightarrow{\boldsymbol{\Delta}}=\overleftrightarrow
{\openone}-{\bf \hat{n}\hat{n}}$ the projector on the transverse
plane defined by the unit vector $\bf\hat{n}$ along the connecting
line between atoms $\alpha$ and $\beta$. This interaction has a
certain strength, depending on the distance between the atoms, via
\be g =\frac{3i}{2k_0r_{\alpha\beta}}e^{ik_0 r_{\alpha\beta}},
\label{g} \e with $k_0=\omega_0/c$, and on the life time of the
excited atomic levels, through $\gamma$. The coupling constant
$|g|\ll 1$ is small in the far-field ($k_0r_{\alpha\beta}\gg 1$),
where near-field interaction terms of order
$(k_0r_{\alpha\beta})^{-2}$ and $(k_0r_{\alpha\beta})^{-3}$ can be
neglected.

Of course, an arbitrary operator $Q$ inserted into Eq.~(\ref{meq})
does not result in a closed differential equation. Our system
consisting of two 4-level atoms leads to $255=4^2\times 4^2-1$
linear coupled equations of motion for the associated expectation
values. We solve them perturbatively to second order in $g$, to
account for the lowest order (double-)scattering processes giving
rise to a nontrivial interference contribution. To keep this in
mind, symbols denoting double scattering intensities and spectra
will carry the subscript 2.

Note that Eq.~(\ref{meq}) describes the evolution of one-point
correlation functions, whereas $G^{(1)}_{\rm ss}(\tau)$ is a {\it
two}-point correlation function. By virtue of the quantum regression
theorem \cite{scully97}, the latter also satisfies Eq.~(\ref{meq}),
with initial conditions extracted from the stationary solution of
(\ref{meq}). In particular, the double scattering counterpart of
$G^{(1)}_{\rm ss}(0)$ is nothing but the stationary average
backscattered light intensity which will be referred to as $I^{\rm
tot}_2$. There is an obvious relation between $I^{\rm tot}_2$ and
$S_2(\nu)$ [obtained by expanding (\ref{G_tau}) to second order, in
(\ref{sp_fixed})]: \be I^{\rm tot}_2=\int_{-\infty}^{\infty}d\nu
S_2(\nu). \label{I_vs_S} \e

The total CBS intensity at backscattering direction can be
decomposed in a sum of two terms, \be I^{\rm tot}_2=L^{\rm
tot}_2+C^{\rm tot}_2, \label{Itot2} \e where $C^{\rm tot}_2\equiv
C^{\rm tot}_2(\theta=0)$ (i.e., ${\bf k}=-{\bf k}_L$), and \beq
C^{\rm
tot}_2(\theta)&=&2\re\la\la\sigma^1_{21}\sigma^2_{12}\ra^{[2]}_{\rm
  ss}e^{i\vk\cdot{\bf
r}_{12}}\ra_{\rm conf}\, ,\label{Cterm}\\
L^{\rm
tot}_2&=&\la\la\sigma_{22}^1\ra^{[2]}_{\rm
  ss}+\la\sigma_{22}^2\ra^{[2]}\ra_{\rm ss}\ra_{\rm
conf}\, ,\label{Lterm} \eq are the so-called ``crossed" ($C^{\rm
tot}_2(\theta=0)$) and ``ladder" ($L^{\rm tot}_2$) terms,
respectively. Using these, we can derive the standard measure of the
phase coherence between the counterpropagating amplitudes that
contribute to CBS -- the enhancement factor \be
\alpha=1+\frac{C^{\rm tot}_2}{L^{\rm tot}_2}. \label{efactor} \e For
perfect two-wave interference, $\alpha=2$. In general, the total
backscattered light intensity has elastic and inelastic components,
\be I^{\rm tot}_2=I^{\rm el}_2+I^{\rm inel}_2, \e with the elastic
component given by products of the expectation values of the atomic
dipoles, \be I^{\rm
el}_2=\sum_{\alpha,\beta=1}^2\la\la\sigma^\alpha_{21}\ra^{[1]}_{\rm
ss}\la\sigma^\beta_{12}\ra^{[1]}_{\rm ss} e^{-i{\bf k}_L\cdot{\bf
r}_{\alpha\beta}}\ra_{\rm conf}. \label{el_part} \e For
$\alpha=\beta$ we obtain the elastic ladder term $L^{\rm el}_2$, and
for $\alpha\neq\beta$ the elastic crossed term $C^{\rm el}_2$. Given
$I^{\rm tot}_2$ and $I^{\rm el}_2$, also the fluctuating part of the
dipole correlation functions defining $I^{\rm inel}_2$ is
determined.

A detailed derivation of the stationary CBS intensity, together with
analytical results for $\delta=0$, can be found in Appendix A.

\section{CBS spectrum}
\label{sect:results} With the above premises, we can now proceed to
the detailed analysis of the CBS spectrum, the central object of
this paper. A detailed derivation of the CBS spectrum from the
solution of Eq.~(\ref{meq}) is given in the Appendix B, while the
physical content thereof will be discussed in the following
sections.
\subsection{Elastic spectrum}
In the saturation regime, the double scattering spectrum of CBS has
elastic and inelastic parts. The elastic spectrum in the
backscattering direction reads \be \tilde{I}^{\rm el}_2(\nu)=I^{\rm
el}_2\delta(\nu), \label{Sp_el} \e where $\delta(\nu)$ is Dirac's
delta-function, and $I^{\rm el}_2=L^{\rm el}_2+C^{\rm el}_2$, with
the ladder and crossed contributions \cite{shatokhin06} \be L^{\rm
el}_2=C^{\rm
el}_2=\frac{2|\bar{g}|^2}{15}\frac{1}{1+(\delta/\gamma)^2}\frac{s}{(1+s)^4},
\label{el_int}\e where $\bar{g}$ is the configuration averaged value
of $g$. A detailed interpretation of the elastic intensity was given
in previous work \cite{shatokhin05,shatokhin06}. Here, we will focus
on the \ldots
\subsection{Inelastic spectrum}
\subsubsection{Normalization}
\label{sec:norm}
\begin{figure}
\includegraphics[width=8cm]{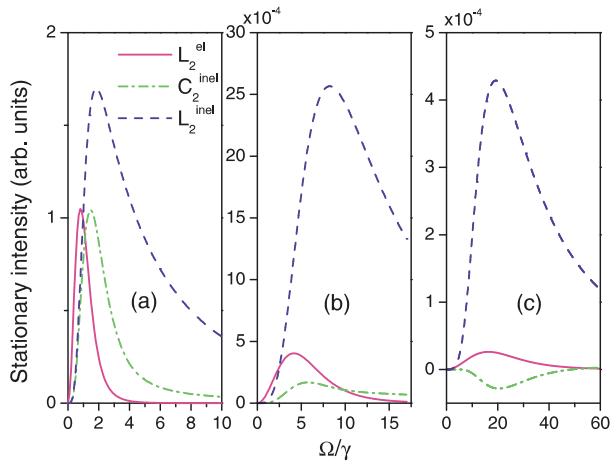}
\caption{(Color online) Elastic ladder, $L_2^{\rm el}$ (solid),
inelastic ladder, $L_2^{\rm inel}$ (dashed), and inelastic crossed,
$C_2^{\rm inel}$, (dashed-dotted) CBS contributions vs. Rabi
frequency $\Omega$, for different values of the detuning $\delta$:
$\delta=$ (a) $0$; (b) $5\gamma$; (c) $20\gamma$.}
\label{fig:normalization}
\end{figure}
Since we are interested in the spectral properties of CBS, we will
omit all frequency-independent prefactors. This is naturally
achieved when considering normalized expressions. We choose the
stationary inelastic ladder contribution $L^{\rm inel}_2$ as
normalization factor, such that the integrals of $\tilde{L}^{\rm
inel}_2(\nu)/L^{\rm inel}_2$ and $\tilde{C}^{\rm inel}_2(\nu)/L^{\rm
inel}_2$ over $\nu$ yield unity and $C^{\rm inel}_2/L^{\rm inel}_2$,
respectively. In the deep saturation regime, the value of the latter
integral tends to the asymptotic value of the interference contrast
of CBS, $\alpha_\infty(\delta)-1$. Furthermore, for arbitrary
saturation, the areas under the peaks of the normalized ladder
spectrum give the relative probabilities of the corresponding
inelastic processes.

The normalization itself depends on the parameters of the driving
field. In Fig.~\ref{fig:normalization}, we present several examples
of the elastic and inelastic intensities, as functions of $\Omega$,
and for different values of the detuning $\delta$. This figure
shows, in particular, that, for $\delta=20\gamma$, the inelastic
crossed term is negative, for a range of Rabi frequencies, with a
minimum at $\Omega\simeq 20\gamma=\delta$ (see
Fig.~\ref{fig:normalization}(c)). Furthermore, note that, since
$C_2^{\rm el}=L_2^{\rm el}$, the total crossed term $C_2^{\rm
el}+C_2^{\rm inel}<0$ around $\Omega=20\gamma$. The values of
$C_2^{\rm el}=L_2^{\rm el}$, $C_2^{\rm inel}$, and $L_2^{\rm inel}$
at $\Omega=20\gamma$ and $\delta=20\gamma$ are $2.45\times 10^{-5}$,
$-2.82\times 10^{-5}$, and $4.27\times 10^{-4}$, respectively, what
implies $C_2^{\rm inel}/L_2^{\rm inel}=-0.066$, and $\alpha=0.991$.
The negativity of the total crossed term thus results in
anti-enhancement (i.e., an enhancement factor $\alpha<1$) and was
reported previously \cite{shatokhin06}. It is one of the purposes of
the present paper to identify the physical origin thereof.

\subsubsection{Weakly inelastic scattering}
\label{sec:weakly_inel} 
\begin{figure}
\includegraphics[width=8cm]{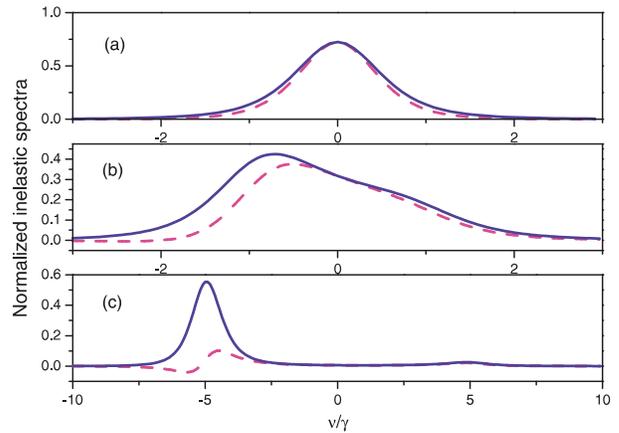}
\caption{(Color online) Normalized inelastic spectra of the ladder,
$\tilde{L}_2^{\rm inel}(\nu)$ (solid), and crossed,
$\tilde{C}_2^{\rm inel}(\nu)$ (dashed), terms at $\Omega=0.1\gamma$.
$\delta=$ (a) $0$; (b) $\gamma$; (c) $5\gamma$.}
\label{fig:N_S_Small_E}
\end{figure}

In the regime of small Rabi frequencies
$\Omega\ll\gamma$, it is the lowest-order inelastic processes --
two-photon processes, proportional to $\Omega^2$ -- that contribute
to the inelastic spectrum. Consequently, the emerging spectral
features of the ladder and crossed spectra derived with the aid of
the master equation approach (see Fig.~\ref{fig:N_S_Small_E}) can be
interpreted on the basis of two-photon amplitudes derived in
\cite{wellens04}. There it was shown that, in the weakly inelastic
regime, one of the two atoms must scatter inelastically.
Furthermore, for direct and (time-)reversed amplitudes (which have
to interfere constructively to create the CBS signal), the
inelastically scattering atom must be the same. Then it turns out
that, although initial and final frequencies of the scattered
photons are the same, the intermediate frequencies differ, leading
to the non-reciprocity of the interfering amplitudes, and a decrease
of the CBS enhancement factor. The non-reciprocity argument was also
successfully applied to interpret a dispersive resonance of the
crossed term's spectrum in the weakly inelastic regime
\cite{gremaud06}.
\begin{figure}
\includegraphics[width=8cm]{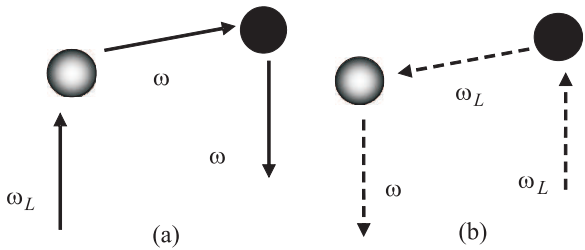}
\caption{Direct (a) and (time-)reversed (b) scattering
paths/amplitudes, giving rise to the CBS signal in the weakly
inelastic scattering regime. Grey-shaded and black spots represent
inelastically and elastically scattering scattering atoms,
respectively. The amplitudes spell out the transformation of ingoing
frequencies $\omega_L$ into outgoing frequencies $\omega$. These
interfering amplitudes are non-reciprocal, since the frequencies of
the intermediate photons for direct and reversed paths are
different.} \label{fig:2photon_diags}
\end{figure}

Figure \ref{fig:2photon_diags} shows direct (a) and reversed (d)
processes contributing to the CBS signal. Without loss of
generality, we assume that the left atom is scattering
inelastically. The direct, $E_1$, and reversed, $E_2$, scattering
amplitudes can be derived as \cite{wellens04}\beq E_1&=&-\frac{
e^{-i\phi/2}(\gamma+i\delta)}{(\gamma+i(\delta-\nu))(\gamma+i(\delta+\nu))^2},
\label{direct_w}\\
E_2&=&-\frac{e^{i\phi/2}}{(\gamma+i\delta)^2+\nu^2},
\label{reversed_w}\eq where $\phi$ is a phase which depends on the
geometric configuration, and should be taken zero in the
backscattering direction. From Eqs.~(\ref{direct_w}) and
(\ref{reversed_w}), we obtain following expressions for the ladder
and crossed spectra: \beq \tilde{L}_2^{\rm
inel}(\nu)&=&\frac{2(\gamma^2+\delta^2)+2\delta\nu+\nu^2}
{(\gamma^2+(\delta-\nu)^2)(\gamma^2+(\delta+\nu)^2)^2},\label{ladd_w}\\
\tilde{C}_2^{\rm inel}(\nu)&=&\frac{2(\gamma^2+\delta(\delta+\nu))}
{(\gamma^2+(\delta-\nu)^2)(\gamma^2+(\delta+\nu)^2)^2}.\label{cross_w}
\eq Expressions (\ref{ladd_w}) and (\ref{cross_w}) precisely
reproduce the spectral lineshape of the ladder and crossed spectra
of Fig.~\ref{fig:N_S_Small_E}, and, up to the prefactor
$\Omega^4/(8\pi\gamma)$, coincide with the respective results of the
master equation approach at vanishing detuning $\delta=0$ (see
Eqs.~(\ref{LCina0}) and (\ref{LCina}) in Appendix B). They also
allow for a transparent interpretation of the CBS spectra for
arbitrary detuning $\delta$.

\begin{figure}
\includegraphics[width=8cm]{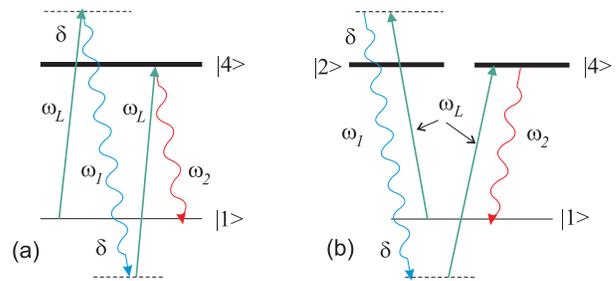}
\caption{Two-photon processes describing the laser frequency
transformation on the left atom in Fig.~\ref{fig:2photon_diags}. (a)
corresponds to Fig.~\ref{fig:2photon_diags}(a): Two laser photons
are converted into two scattered photons such that
$\omega_1\simeq\omega_0+2\delta$ and $\omega_2\simeq\omega_0$.
Either one of these photons then travels to the other atom. (b)
corresponds to Fig.~\ref{fig:2photon_diags}(b) in case
$\omega\simeq\omega_1$: One laser photon and one elastically
scattered photon give rise to the CBS photon at
$\omega\simeq\omega_1$ and the undetected fluorescence photon at
$\omega_2\simeq\omega_0$. The roles of the photons at $\omega_1$ and
$\omega_2$ can be interchanged by flipping their polarizations.}
\label{fig:2photon}
\end{figure}

By inspecting the denominators of Eqs.~(\ref{ladd_w}) and
(\ref{cross_w}), we see that, at $\delta=0$, both $\tilde{L}_2^{\rm
inel}(\nu)$ and $\tilde{C}_2^{\rm inel}(\nu)$ must exhibit a single
peak, at $\nu=0$, as is the case in Fig.~\ref{fig:N_S_Small_E}(a).
At $\delta\neq 0$, there must be two resonances at
$\nu\simeq\pm\delta$, with the more pronounced one at
$\nu\simeq-\delta$ (since the respective term is squared) (see
Fig.~\ref{fig:N_S_Small_E}(b,c)). The physical reason for this
behavior is easy to understand by recalling that upon inelastic
scattering on the (left, in Fig.~\ref{fig:2photon_diags}) atom, the
re-emitted photon frequency $\omega$ can be either
$\omega_L-\delta=\omega_0$ or $\omega_L+\delta=\omega_0+2\delta$, as
spelled out by the level diagram in Fig.~\ref{fig:2photon}.
The former, equal to the atomic resonance frequency, corresponds to
$\nu\simeq-\delta$, and acquires a large scattering cross-section,
while the latter corresponds to $\nu\simeq+\delta$, and has a
diminished scattering cross-section since detuned by $2\delta$ from
the atomic transition frequency.

Another peculiarity of the CBS spectra which clearly stands out in
Fig.~\ref{fig:N_S_Small_E}(c) as well as in Eq.~(\ref{cross_w}) is a
dispersive lineshape of the crossed spectrum around
$\nu\simeq-\delta$. In other words, as the signal frequency passes
from $\omega>\omega_0$ to $\omega<\omega_0$, the interference
character between the interference paths (a) and (b) of
Fig.~\ref{fig:2photon_diags} turns from constructive to destructive.
This is due to the continuous $\omega$-dependence of the phase shift
between the direct (\ref{direct_w}) and reversed (\ref{reversed_w})
scattering amplitudes, passing through $\pi/2$ at
$\omega\simeq\omega_0$.

\subsubsection{Strongly inelastic scattering}
\label{sec:strongly_inelastic} 
\begin{figure}
\includegraphics[width=9cm]{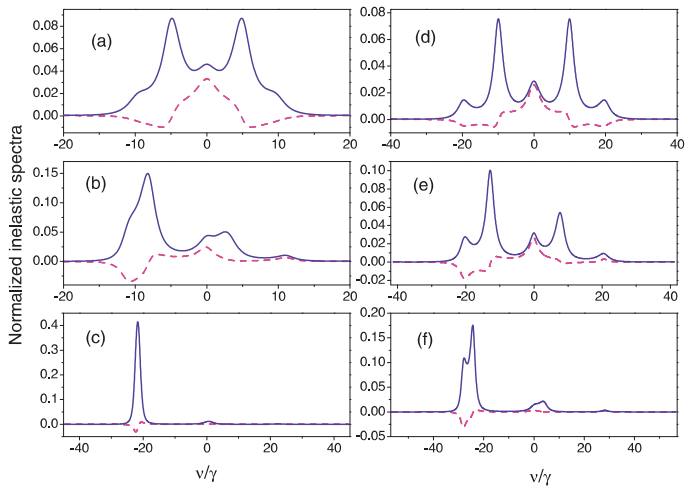}
\caption{(Color online) Normalized inelastic spectra of the ladder,
$\tilde{L}_2^{\rm inel}$, (solid) and crossed, $\tilde{C}_2^{\rm
inel}$, (dashed line) terms. Left panel ((a)-(c)):
$\Omega=10\gamma$; right panel ((d)-(f)): $\Omega=20\gamma$. The
detuning increases from top to bottom: $\delta=0$ (top); $5\gamma$
(middle); $20\gamma$ (bottom).} \label{fig:N_S_Big_E}
\end{figure}

As the Rabi frequency $\Omega$
increases to values $\Omega>\gamma$, scattering processes of higher
than second order contribute for the individual atoms, resulting in
the emission of the resonance fluorescence Mollow triplet
\cite{mollow69}. Correspondingly, the CBS spectra become more
complicated at intense driving. Figure \ref{fig:N_S_Big_E} presents
examples of normalized ladder and crossed spectra for two fixed
values of the Rabi frequency, $\Omega=10\gamma$
(Fig.~\ref{fig:N_S_Big_E}(a-c)), and $\Omega=20\gamma$
(Fig.~\ref{fig:N_S_Big_E}(d-f)).

At exact resonance, the spectra remain symmetric, with signatures of
five resonances at $\nu=\pm \Omega$, $\nu=\pm\Omega/2$, and $\nu=0$.
Crossed spectra take negative values in a frequencies range which
spares out the central, positive, peak. With increasing $\delta$,
some of the resonances approach each other (observe, e. g., the
change of position of the two left-most resonances of the ladder
term from Fig.~\ref{fig:N_S_Big_E}(a) to (b), and from (d) to (e)),
and eventually merge (see Fig.~\ref{fig:N_S_Big_E}(f)).
Fig.~\ref{fig:N_S_Big_E}(c) is qualitatively reminiscent of
Fig.~\ref{fig:N_S_Small_E}(c), though additionally garnished with a
signal at $\nu\simeq 0$ in both the ladder and crossed spectra. This
additional resonance stems from the central peak of the Mollow
triplet, emerging in three-photon scattering processes on one atom.
In contrast, Fig.~\ref{fig:N_S_Small_E}(c) corresponds to a much
weaker saturation parameter ($s=2\times 10^{-4}$) where three-photon
processes are negligible, and hence the central Mollow peak is not visible..

A notable feature of the spectra of Fig.~\ref{fig:N_S_Big_E} is
that, for $\delta=20\gamma$ ((c) and (f)), the overall inelastic
crossed term becomes negative, i.e.,
$\int_{-\infty}^{\infty}d\nu\tilde{C}_2^{\rm inel}(\nu)/L_2^{\rm
inel}=-0.029$ (for $\Omega=10\gamma$), and $-0.066$ (for
$\Omega=20\gamma$), respectively. The negative value of the
normalized crossed term in the latter case dominates the positive
ratio $C_2^{\rm el}/L_2^{\rm inel}=0.057$ which can be extracted
from Fig.~\ref{fig:normalization}(c)
for $\Omega=20\gamma$.
Therefore, the enhancement
factor $\alpha=1+(C_2^{\rm el}+C_2^{\rm inel})/(L_2^{\rm
el}+L_2^{\rm inel})$ is equal to $0.991<1$, in accordance with
\cite{shatokhin06} and Sec.~\ref{sec:norm}. It is now clear that the
anti-enhancement comes from destructive self-interference of
inelastically scattered photons around
$\nu=-28\gamma\simeq-\sqrt{\Omega^2+\delta^2}$, with
$\Omega=\delta=20\gamma$. As seen in Fig.~\ref{fig:N_S_Big_E}(f), at
these parameter values, the two resonances at which the crossed
spectrum is negative are almost merged. Hence, the destructively
interfering processes responsible for each distinct resonance become
indistinguishable, and interfere with each other. This interference
additionally broadens the negative area of the crossed spectrum,
leading to anti-enhancement. As the detuning becomes much larger
than the Rabi frequency, only the dispersive resonance of the
crossed spectrum survives, leading to a picture qualitatively
similar to that of Figs.~\ref{fig:N_S_Small_E}(c),
\ref{fig:N_S_Big_E}(c). The position of this resonance corresponds
to the resonance frequency of an ac-Stark shifted CBS transition,
and will be specified in the next subsection.

\subsubsection{Limit of well-separated spectral lines}
\label{sec:well_resolved} Although the positions of the resonances
of the CBS spectrum could be guessed from the results presented in
the previous section, the spectral line shape for the ladder and
crossed terms is not quite clear, since not yet fully resolved at
the Rabi frequency considered. To reach a fully transparent picture,
we will now address the limit of well-separated spectral lines, at
$\Omega\gg\gamma$.
\begin{figure}
\includegraphics[width=8.5cm]{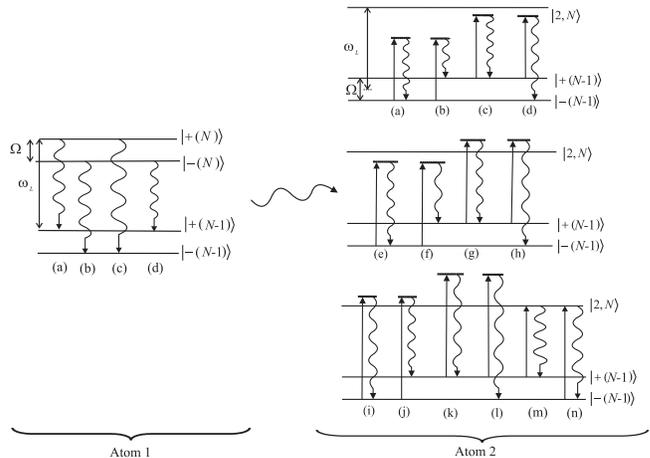}
\caption{ Single (left) and double (right) scattering processes.
Only the laser-driven and the CBS transition are depicted,
respectively. Radiative processes (a) and (b) on the left give rise
to the Rayleigh peak of the Mollow triplet with linewidth $\gamma$
centered at $\nu=0$. (c)
 and (d) give rise to the sidebands at $\nu=\Omega$ and $\nu=-\Omega$,
 respectively, with linewidths $3\gamma/2$.
 Photons emitted by atom 1 (left) propagate to atom 2 (right), and are
 scattered from either one of the dressed states of the
 energy manifold with $N-1$ laser photons. The diagrams on the right describe
 the scattering of: (a)-(d) a sideband
 photon centered at $\nu=-\Omega$; (e)-(h) a
 photon from the central peak; (i)-(l) a sideband photon at $\nu=\Omega$;
 (m), (n) resonant photons at $\nu=\pm\Omega/2$.}
\label{fig:diagRes}
\end{figure}
\paragraph{Exact resonance.}
Let us start with a reminder of the level structure of the pumped
(by the driving laser) and probed (by the scattered photon)
transitions of both atoms. In an intense laser field with
$\Omega\gg\gamma$, the pumped transition $|1\ra\leftrightarrow|4\ra$
is strongly coupled to the laser mode. In this case, it is
instructive to treat the latter as a quantum system
\cite{shirley65,cohen_tannoudji}. The eigenstates of the laser-atom
interaction Hamiltonian are the dressed states $|\pm(N)\ra_\alpha$.
For $\delta=0$, they read \be
|\pm(N)\ra_\alpha=\frac{1}{\sqrt{2}}(|1,N+1\ra_\alpha\pm e^{i{\bf
k}\cdot{\bf r}_\alpha}|4,N\ra_\alpha), \label{dr_st} \e where $N$
and $N+1$ refer to the number of photons in the laser mode, and
$\alpha$ labels the atoms. Inasmuch as the dressed states represent
superpositions of the ground and {\it excited} atomic states of the
laser-driven transition, they have widths equal to $\gamma$ for the
resonant driving \cite{cohen_tannoudji}. Spontaneous transitions
from the dressed states manifold $\{|\pm(N)\ra_\alpha\}$ to
$\{|\pm(N-1)\ra_\alpha\}$ lead to emission of the fluorescence
spectrum centered at frequencies $\omega_L-\Omega$, $\omega_L$, and
$\omega_L+\Omega$ known as the Mollow triplet \cite{mollow69} (see
the left of Fig.~\ref{fig:diagRes}). The width of the central peak
is defined by the decay rate of the dressed state's populations, and
is equal to $2\gamma$, while the width of the sidebands is
determined by the decay rate of the coherences between the dressed
states, and is equal to $3\gamma$ \cite{cohen_tannoudji}.

On the other hand, the level $|2\ra$ is not affected by the laser
field. As a result, the probed transition
$|1\ra\leftrightarrow|2\ra$ of both atoms, giving rise to CBS
photons, is affected by the laser field only via the state $|1\ra$.
For that reason, new resonance frequencies of the probed transition
emerge at $\omega_L\pm\Omega/2$ (the right of
Fig.~\ref{fig:diagRes}).

When the Mollow triplet emitted by one atom is incident on another
atom, it is scattered on the internal structure of the latter.
Relevant scattering processes that can take place are depicted on
the right of Fig.~\ref{fig:diagRes}. Each photon can be scattered
either elastically or undergo Raman-Stokes or -anti-Stokes
(multiphoton) transitions \cite{Boyd} (which lead to a frequency
change by $-\Omega$ or $\Omega$, respectively), which conserve
energy and angular momentum. It follows that the CBS spectrum must
have resonances at $\nu=\pm 2\Omega$, $\nu=\pm\Omega$, and $\nu=0$.
The diagramms describing the emission of CBS photons at these
frequencies are (a)-(l) on the right of Fig.~\ref{fig:diagRes}.

However, there are two more diagrams, (m) and (n), that apparently
do not fit into the scheme just described. On these diagrams,
photons with frequencies $\nu=\pm\Omega/2$ are resonantly scattered
to give rise to an additional doublet in the CBS spectrum. This is
nothing but an Autler-Townes doublet \cite{townes55,mollow72b}: One
should bear in mind that a Lorentzian distribution has long tails,
and thus emission of photons with frequencies very different from
the central frequencies is not impossible, though with very small
probabilities. For instance, two laser photons may be transformed
into two fluorescence photons with frequencies $\omega_L+\Omega/2$
and $\omega_L-\Omega/2$, with each of these photons appearing as a
result of quantum interference between the Rayleigh and
Raman-anti-Stokes or -Stokes transitions, respectively
\cite{shatokhin01}.

What makes the frequencies $\omega_L\pm\Omega/2$ special in our
present problem is that both (dressed) atoms have two transitions
{\it exactly in resonance} with these frequencies. Therefore,
although the probability of the creation of a pair of photons with
frequencies $\omega_L\pm\Omega/2$ is relatively small, the
probability of their {\it double} scattering is high, due to a large
resonant scattering cross-section on the transition
$|1\ra\leftrightarrow|2\ra$ modified by the ac-Stark effect. The
overall effect finally acquires the same strength as the elastic
(though non-resonant) scattering of photons which are emitted into
the dominant frequency components of the Mollow triplet, by the
first atom.

\begin{figure}
\includegraphics[width=8.5cm]{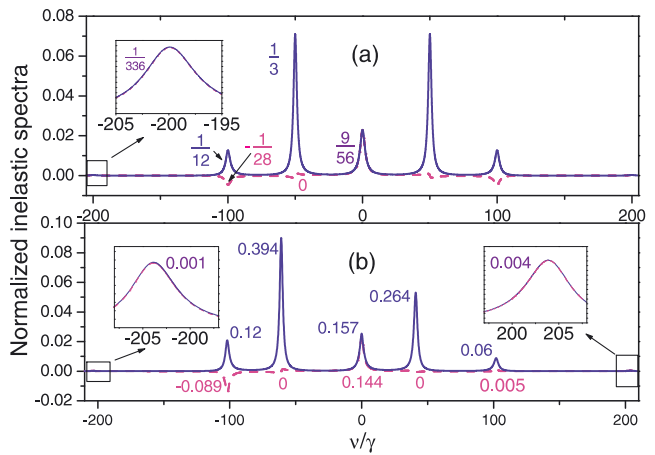}
\caption{(Color online) Normalized inelastic spectra of the ladder,
$\tilde{L}_2^{\rm inel}(\nu)$ (solid), and crossed,
$\tilde{C}_2^{\rm inel}$ (dashed), terms, in the limit of well
separated spectral lines, at $\Omega=100\gamma$. $\delta=$ (a) $0$;
(b) $20\gamma$. The numbers near the resonances indicate their
areas, such that the overall areas of the ladder and crossed terms
give unity and $C_2^{\rm inel}/L_2^{\rm inel}$, respectively.}
\label{fig:N_S_Limit_E}
\end{figure}

The results of our calculation are presented in
Fig.~\ref{fig:N_S_Limit_E}(a).
Consistently with our analysis, both the ladder and crossed spectra
consist of seven resonances, with the positions precisely at the
predicted frequencies, that is, at $\nu=\pm 2\Omega$,
$\nu=\pm\Omega$, $\nu=\pm \Omega/2$, $\nu=0$. The respective
analytic expressions, (\ref{lad:asymp}) and (\ref{cro:asymp}), can
be found in Appendix B.

Let us now discuss the ladder spectrum in more detail, while
postponing the analysis of the spectrum of the interference
contribution for the next subsection. In the ladder spectrum, three
of its resonances, at $\nu=0$ and $\nu=\pm\Omega$, represent sums of
two Lorentzians with different widths. The remaining ones are simple
Lorentzians. In order to understand this structure quantitatively,
let us reinspect the diagrams on the right of
Fig.~\ref{fig:diagRes}: We start with the central CBS resonance at
$\nu=0$. Several processes, (d), (e), (g), and (j) on the right of
Fig.~\ref{fig:diagRes} contribute. Diagrams (e) and (g) represent a
photon emitted by the first atom into the inelastic Rayleigh
component (centered at $\omega_L$), elastically scattered on the
dressed states $|+(N-1)\ra$ and $|-(N-1)\ra$. The corresponding
amplitudes add coherently, so that the two scattering processes are
equivalent to elastic scattering by the atomic ground state $|1\ra$.
Since the ground state does not have a linewidth, the resulting
frequency distribution is the same as for the Rayleigh component of
the resonance fluorescence spectrum, that is, the width $2\gamma$ of
the excited atomic levels. In contrast, the processes (d) and (j)
involve sideband photons centered at either $\omega_L+\Omega$ or
$\omega_L-\Omega$ undergo Raman process on $|+(N-1)\ra$ or
$|-(N-1)\ra$, respectively. This leads to an additional broadening
of the frequency distribution of photons, associated with these
transitions, by $3\gamma$. Neither of the processes (d) and (j)
interferes with (e) and (g), because the frequency changes associated
with these transitions are different, and information about the latter
is carried into the environment by the  undetected photons \cite{wellens04}.
As a result, each of the processes (d) and (j)
contributes to the CBS ladder spectrum with a Lorentzian centered at
$\nu=0$, with linewidth $6\gamma$, which adds to the one due to the
Rayleigh transitions.

As for the sidebands at $\nu=\pm\Omega$, there are three processes
contributing to each of them. Since the spectrum is symmetric, let
us analyse the low-frequency sideband. The relevant processes are
(a), (c), and (f) in Fig.~\ref{fig:diagRes}. Processes (a) and (c)
are completely analogous to (e) and (g). That is, they add
coherently, and the resulting linewidth is the same as for sideband
photons of the single-atom spectrum, that is, $3\gamma$. The process
(f) is analogous to the process (j), hence the linewidth of the
respective Lorentzian must be the sum of the linewidths of the
Rayleigh photon and of a dressed state. Thus, we obtain a Lorentzian
with linewidth $5\gamma$ centered at $\omega_L-\Omega$.

We proceed in the same manner with the sidebands at $\nu=\pm
2\Omega$. The respective diagrams for the lower-frequency sideband
are (b) and (l). Since, in both cases, the sideband photons of the
resonance fluorescence spectrum undergo Raman processes, the
linewidth of the associated Lorentzian is $6\gamma$.

Also note that the crossed term has the same lineshape as the ladder
term, at $\nu=\pm 2\Omega$ [see Fig.~\ref{fig:N_S_Limit_E}(a)]. This
peculiarity of the CBS spectrum will be discussed in more detail in
the next subsection. In brief, it is related to the fact that there
is only one direct and its reciprocal process, which contribute to
the CBS spectral lines at either one of these frequencies.
Therefore, these inelastic photons (self-)interfere perfectly well.
Although the contribution of the processes at $\nu=\pm 2\Omega$ is
very small, it is non-negligible. The numbers for the integral
contributions of the ladder and crossed terms to the resonances in
Fig.~\ref{fig:N_S_Limit_E}(a) (these numbers can be easily extracted
from Eqs.~(\ref{lad:asymp}) and (\ref{cro:asymp})) yield precisely
the asymptotic value of the enhancement factor $\alpha_\infty=23/21$
derived in \cite{shatokhin06}. Without the contributions from the
outer sidebands, one would underestimate $\alpha_\infty$ by
approximately 6\%.

Finally, let us consider the doublet at $\omega_L\pm\Omega/2$. The
respective diagrams in Fig.~\ref{fig:diagRes}, (m) and (n), describe
the resonant scattering of photons on the transitions
$|+(N-1)\ra\rightarrow|2\ra$ and $|-(N-1)\ra\rightarrow|2\ra$. Since
the incoming photons do not originate from the (Mollow) peaks of the
resonance fluorescence spectrum emitted by the first atom, the
corresponding linewidths are only defined by the linewidths of the
`filtering' transitions of the second atom. These linewidths
$3\gamma$ are a sum of the width $2\gamma$ of the excited level
$|2\ra$, and the linewidth $\gamma$ of either of the dressed states.

\paragraph{Detuned case.}
For $\delta\neq 0$, the major part of the above analysis for
$\delta=0$ is still valid. Though some modifications are needed, in
order to explain why some resonances creep towards the other ones;
their weights are redistributed, and even the interference character
of some of them is changed (see Fig.~\ref{fig:N_S_Limit_E}(b)).
Here, we will comment on their linewidths and positions.

To derive the new linewidths of the ladder term, one needs to
account for the dependence of linewidths of the resonance
fluorescence spectrum of the first atom on the detuning
\cite{mollow69}. With that and Fig.~\ref{fig:diagRes}, one easily
obtains the linewidths of the CBS ladder spectrum, at finite
$\delta$.

It is easy to show that the new resonance frequencies of the CBS
transition $|1\ra\leftrightarrow|2\ra$, modified by the ac-Stark
effect, are $\omega_0\pm(\Omega^{\prime}\pm\delta)/2$, where
$\Omega^\prime=\sqrt{\Omega^2+\delta^2}$ is the modified Rabi
frequency \cite{mollow69}. It follows that for the detuned case,
seven CBS resonances manifest at $\nu=\pm 2\Omega^{\prime}$
$\nu=\pm\Omega^\prime$, $\nu=\pm(\Omega^\prime\mp\delta)/2$, and
$\nu=0$. Therefore, as we increase the detuning, the Autler-Townes
doublet centered at $\nu=-(\Omega^\prime+\delta)/2$ approaches the
sideband at $\nu=-\Omega^\prime$, whereas its counterpart centered
at $\nu=(\Omega^\prime-\delta)/2$ approaches the central resonance,
as evident from Fig.~\ref{fig:N_S_Limit_E}(b).

Since the single-atom resonance fluorescence spectrum is always
symmetric, this also explains why the sideband at
$\nu=-\Omega^\prime$ which is red-shifted with respect to $\nu=0$
gains more weight than its blue-shifted counterpart: It is closer to
the transition frequency
$\omega_0^\prime=\omega_0-(\Omega^\prime-\delta)/2$ as compared to
the distance of the sideband at $\nu=\Omega^\prime$ from
$\omega_0^{\prime\prime}=\omega_0+(\Omega^\prime-\delta)/2$.
Consequently, the associated scattering cross-section is larger.
This also explains a larger value of the peak at
$\nu=-(\Omega^{\prime}+\delta)/2$ than at
$\nu=(\Omega^{\prime}-\delta)/2$, given that, under the detuned
driving, these resonances originate mainly from the two sidebands of
the Mollow triplet. However, when we consider Raman scattering at
$\nu=\pm 2\Omega^\prime$, the scattering cross-section for a photon
at $\omega_L+\Omega^\prime$ depends on its detuning
$(\Omega^\prime+\delta)/2$ from the transition
$|2,N\ra\leftrightarrow|+(N-1)\ra$ [see Fig.~\ref{fig:diagRes}(l)].
Likewise, the cross-section for a photon at $\omega_L-\Omega^\prime$
is determined by its detuning $(3\Omega^\prime-\delta)/2$ from the
transition $|2,N\ra\leftrightarrow|-(N-1)\ra$ [see diagram (b) on
the right of Fig.~\ref{fig:diagRes}]. For $\Omega=100\gamma$ and
$\delta=20\gamma$, the latter detuning is obviously larger, leading
to a redistribution of signal weights in favor of the blue-shifted
sideband [compare the insets of Fig.~\ref{fig:N_S_Limit_E}(b)]. Note
also that this asymmetry does not spoil the perfectness of the
interference of CBS photons around $\nu=\pm 2\Omega^\prime$.

\subsection{Interpretation of the crossed spectrum at $\delta=0$}
\label{sec:crossed} Finally, for an intuitive interpretation of the
interference character of different components of the crossed term's
spectrum, we will rely upon the diagrammatic technique developed in
\cite{wellens04}. We will see that a slightly modified version
thereof in terms of dressed states allows for the explanation of the
destructive interference at $\nu=\pm\Omega$, of the dispersive
resonances at $\nu=\pm\Omega/2$, and of the constructive
interferences at the other resonances.

Let us first note that the interference character of different
resonances of the crossed spectrum is defined by how much, if at
all, the outgoing light frequency is shifted with respect to the
incoming laser frequency $\omega_L$. In the secular limit
$\Omega\gg\gamma$, these shifts are of the order of $\Omega$, with
the widths of the shifts' distribution of the order of $\gamma$.
Frequency shifts give rise to phase shifts between the interfering
amplitudes. Taking into account the continuous $\omega$-dependence
of the amplitudes' phases, we will ignore the frequency changes of
the order of $\gamma$, and refer to the respective scattering
processes as `quasi-elastic'. We will label `inelastic' the
processes upon which the frequency changes by the quantity of the
order of $\Omega$.

\begin{figure}
\includegraphics[width=7.5cm]{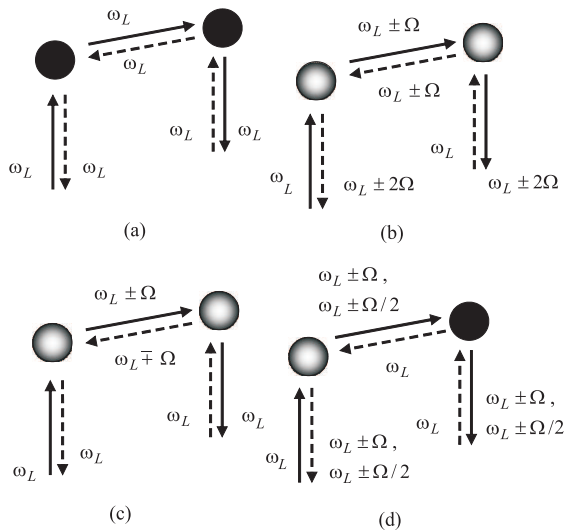}
\caption{Qualitative interpretation of the interference character.
Grey-shaded and black spots represent atoms scattering inelastically
and quasi-elastically, respectively. Solid and dashed arrows
correspond to direct and reversed scattering amplitudes,
respectively. In examples (a) and (b), the direct and the reversed
amplitudes are reciprocal, whereas they are not in cases (c) and
(d).} \label{fig:diags2}
\end{figure}

With these preliminaries, we will use diagrams similar to that of
Fig.~\ref{fig:2photon_diags}, to depict direct and reversed
scattering amplitudes for the scattering processes contributing to
the CBS spectrum, in the limit of well-separated spectral lines (see
Fig.~\ref{fig:N_S_Limit_E}). These diagrams, shown in
Fig.~\ref{fig:diags2}, split into three categories: (a), both atoms
scatter quasi-elastically; (b), (d), both atoms scatter
inelastically; (c), one of the two atoms scatters inelastically.
When both atoms scatter quasi-elastically
[Fig.~\ref{fig:diags2}(a)], or inelastically -- into the outer
sidebands [Fig.~\ref{fig:diags2}(b)], there is a reciprocal,
reversed process, such that the interference is perfect, as indeed
observed for peaks at $\nu\simeq0$ and $\nu\simeq\pm 2\Omega$ of
Fig.~\ref{fig:N_S_Limit_E}(a), since there, the crossed and the
ladder contributions are indistinguishable.
However, note that, since each inelastic event induces the same
change of frequency for both reversed processes, the intermediate
photons ($\omega_L\pm\Omega$) have opposite detuning from the laser frequency in Fig.~\ref{fig:diags2}(c).
In the case $\delta=0$, those two frequencies are equally far detuned from
the atomic resonance, and, as evident from Fig.~\ref{fig:N_S_Limit_E}(a),
the equality between the reversed amplitudes remains preserved.
Only for non-vanishing detuning, the interference contrast at $\nu=0$ is
slightly reduced, see Fig.~\ref{fig:N_S_Limit_E}(b), where the
weights of the crossed and ladder contribution ($0.144$ and $0.157$)
are different.

Now let us address diagram (d) of Fig.~\ref{fig:diags2} which is
reminiscent of Fig.~\ref{fig:2photon_diags}. This diagram shows
non-reciprocal interfering amplitudes, which, depending on the
frequency shifts between the intermediate photons for direct and
reversed paths, can describe either the dispersive features (which
corresponds to $\nu=\pm\Omega/2$) of the crossed term's spectrum, or
the destructive interference thereof (at $\nu=\pm\Omega$). In order
to see this, we will construct explicit expressions for the direct
and reversed scattering amplitudes. Although both atoms are driven
by a powerful laser field, it is sufficient to consider the {\it
two-photon} scattering amplitudes, as in
Fig.~\ref{fig:2photon_diags}, provided that the ac-Stark shifts of
the CBS transition $|2\ra\leftrightarrow|1\ra$ as well as of the
laser-driven transition $|4\ra\leftrightarrow|1\ra$ are properly
accounted for. In other words, for the CBS transition, instead of
considering the atomic resonance frequency $\omega_0$, one should
consider two resonance frequencies
$\omega_0^\prime=\omega_0-\Omega/2$ and
$\omega_0^{\prime\prime}=\omega_0+\Omega/2$. On the other hand,
$\omega_0$ remains the resonance frequency for the laser-driven
transition. It should be stressed that our present treatment is very
schematic and does not pretend to be a full explanation of the
interference character in the saturation regime. Rather, we aim here
at a qualitative understanding of the interference effect as
observed in Fig.~\ref{fig:N_S_Limit_E}.

In order to construct the direct and reversed amplitudes, $E_1$ and
$E_2$, respectively, we employ the symmetry of the CBS spectrum at
$\delta=0$ with respect to a change of the sign of $\nu$, and
consider only the interference character of the two resonances, at
$\nu_*\simeq\Omega,\, \Omega/2$. In the direct two-photon process,
photons with frequencies $\omega_L+\nu_*$ and $\omega_L-\nu_*$ are
scattered by the first atom. We assume that the detected photon with
frequency $\omega_*=\omega_L+\nu_*$ is quasi-elastically scattered
on an ac-Stark shifted transition with frequency
$\omega_0^{\prime\prime}$ of the second atom, since $\omega_*$ is
closer to the modified transition frequency
$\omega_0^{\prime\prime}$ than to the one with frequency $\omega_0^{\prime}$. For the
reversed amplitude, a laser photon is scattered quasi-elastically by
the second atom. Then, the first atom scatters inelastically a pair
of photons with frequency $\omega_L$ into photons with frequencies $\omega_*$ and
$2\omega_L-\omega_*$, with the detected photon originating from the
Raman-anti-Stokes scattering on the ac-Stark shifted transition with
the resonance frequency $\omega_0^{\prime}$, while the undetected photon
scatters on the laser-driven transition, for which $\omega_0$ is the
resonance frequency.

With the above assumptions, the amplitudes $E_1$ and $E_2$ read \beq
E_1\!&=&\!\Bigl(\frac{1}{\nu_*-i\gamma}\!-\!\frac{1}{\nu_*+
i\gamma}\Bigr)
\frac{e^{-i\phi/2}}{\nu_*-\Omega/2-i\gamma},\label{E1_big_W}\\
E_2\!&=&\!\Bigl(\frac{1}{\nu_*+\Omega/2-i\gamma}
-\!\frac{1}{\nu_*+i\gamma}\Bigr)
\frac{ie^{i\phi/2}}{\gamma}.\label{E2_big_W} \eq

It is easy to check that the phase shift between the non-reciprocal
amplitudes (\ref{E1_big_W}) and (\ref{E2_big_W}) continuously
depends on $\omega_*$ (or $\nu_*$). For $\nu_*\simeq\Omega/2$, in
full analogy with the case studied in Sec.~\ref{sec:weakly_inel}
[since $\nu_*$ corresponds to the (modified) transition frequency],
the interfering amplitudes are phase shifted by approximately
$\pi/2$, and thus describe a dispersive line shape, under slight
variation of $\nu_*$. The same amplitudes, for $\nu_*\simeq\Omega$,
are phase shifted by approximately $\pi$ and therefore interfere
destructively. This is quite analogous to the interference-induced
anti-enhancement of CBS in the regime of elastic scattering, due to
the non-reciprocity of combined Rayleigh and Raman scattering
processes \cite{kupriyanov04}. Both the dispersive line shape and
the anti-enhancement is precisely what is observed in
Fig.~\ref{fig:N_S_Limit_E}(a).

\section{Conclusion}
\label{sec:conclusion} We presented a detailed calculation of the
spectrum of coherent backscattering of light by two identical,
randomly placed atoms, for arbitrary strengths of the laser field.

We saw that the CBS spectrum exhibits its most complicated structure
in the limit of strong driving. The ladder term's spectrum reveals
seven Lorentzian peaks which represent: (i) Mollow triplet photons
(re-)scattered by the second atom, with resonances at
$\omega=\omega_L-\Omega$, $\omega=\omega_L$,
$\omega=\omega_L+\Omega$; (ii) an Autler-Townes doublet with
resonances at $\omega=\omega_L-\Omega/2$ and
$\omega=\omega_L+\Omega/2$; (iii) a doublet originating from the
Raman Stokes scattering of the low-frequency sideband, and from the
Raman anti-Stokes scattering of the high-frequency sideband of the
Mollow triplet, with resonance frequencies $\omega=\omega_L-2\Omega$
and $\omega=\omega_L+2\Omega$, respectively.

The crossed term's spectrum reveals resonances located at the same
frequencies as those of the ladder term. We interpreted the
interference character of different resonances through reciprocity
arguments applied to dressed states. Interference is always
constructive for the central resonance at $\omega=\omega_L$, and for
the outer sidebands at $\omega=\pm 2\Omega$. It changes character in
the vicinity of the Autler-Townes doublet (at
$\omega=\omega_L\pm\Omega/2$, where the crossed term's spectrum is
described by a dispersive curve), and contributes a negative
Lorentzian at $\omega=\omega_L\pm\Omega$.

For the case of the detuned driving, at $|\delta|\simeq\Omega$, one
of the dispersive resonances overlaps with one of the negative
Lorentzians. Interference between the scattering processes
responsible for appearance of these two resonances may lead to CBS
anti-enhancement.

\begin{acknowledgements}
Useful discussions with Dominique Delande, Sergei Kilin, Dmitry
Kupriyanov, Christian Miniatura, Cord M\"uller, Alexander Nizovtsev,
Marlan Scully, Igor Sokolov, and Carlos Viviescas are gratefully
acknowledged.

\end{acknowledgements}

\appendix
\section{CBS intensity and enhancement factor}
\subsection{Derivation procedure}
\label{subsub:der_proc}
The operator master equation (\ref{meq}) leads
to the linear matrix equation
\be
\la\dot{\bf Q}\ra=({\bf A}+{\bf V})\la{\bf Q}\ra+{\bf j}.
\label{matr_eq}
\e
Here, vector ${\bf Q}$ having 255 elements is obtained from the tensor product
${\bf q}_1\otimes{\bf q}_2$, where
\beq
{\bf q}&=&(\openone/2,\mu_{1}/2,\mu_{2}/2,\mu_{3}/2,\sigma_{14},\sigma_{41},
\sigma_{13},\sigma_{31},\sigma_{12},\n\\
&&\sigma_{21},\sigma_{34},\sigma_{43},\sigma_{42},\sigma_{24},
\sigma_{32},\sigma_{23})^T \label{defq} \eq
is a vector whose elements comprise the complete
orthonormal basis set of operators for
a four-level quantum system, and indices `1' and `2' number atoms.
In Eq.~(\ref{defq}),
\beml
\beq
\openone\!\!\!&=&\!\!\!\sigma_{11}+\sigma_{22}+\sigma_{33}+\sigma_{44},\\
\mu_1\!\!\!&=&\!\!\!\sigma_{22}-\sigma_{33}+\sigma_{44}-\sigma_{11},\\
\mu_2\!\!\!&=&\!\!\!\sigma_{22}-\sigma_{33}-\sigma_{44}+\sigma_{11},\\
\mu_3\!\!\!&=&\!\!\!\sigma_{22}+\sigma_{33}-\sigma_{44}-\sigma_{11}.
\eq \label{defs_diag} \eml In order to describe a mapping of the
elements of the two $16\times 16$ vectors onto the vector with 255
elements, it is convenient to numerate the basis operators starting
from `0'. We will denote the $i$th element of a vector ${\bf a}$ by
$[{\bf a}]_i$. For instance, $[{\bf q}_1]_0=\openone_1/2$. With
these rules, we map element $\la [{\bf q}_1]_l\otimes [{\bf
q}_2]_m\ra$ onto $\la[{\bf Q}]_n\ra$, where $n=16l+m$, $0\leq
l,m\leq 15$. After we exclude the first element of the tensor
product (it corresponds to $l=m=0$), we obtain that $n$ runs from 1
to 255.  Then, the matrices ${\bf A}$, ${\bf V}$, and ${\bf j}$ are
generated by inserting 255 elements of the vector ${\bf Q}$ to
Eqs.~(\ref{L_a}), (\ref{L_ab}) and performing the quantum mechanical
averaging: \beml \beq \la({\cal L}_\alpha+{\cal L}_\beta)[{\bf
Q}]_n\ra\!\!\!&=&\!\!\! \sum_{m=1}^{255}A_{nm}\la[{\bf
Q}]_m\ra+[{\bf
j}]_n,\\
\la({\cal L}_{\alpha\beta}+{\cal L}_{\beta\alpha})[{\bf Q}]_n\ra\!\!\!&=&\!\!\!\sum_{m=1}^{255}V_{nm}\la [{\bf 
Q}]_m\ra. \eq \eml From the Laplace transform solution to
Eq.~(\ref{matr_eq}), $\la\tilde{\bf Q}(z)\ra=(z\openone_{255}- {\bf
A}-{\bf V})^{-1}(\la{\bf Q}(0)\ra+z^{-1}{\bf j})$, where $z$ is
defined in the main text after Eq.~(\ref{sp_fixed}), and
$\openone_{255}$ is the unit $255\times 255$ matrix, we extract the
steady state solution $\la{\bf Q}\ra_{\rm ss}=\lim_{z\to
0}z\la\tilde{\bf Q}(z)\ra$.

The double scattering contribution corresponds to the perturbative
expansion of $\la{\bf Q}\ra_{\rm ss}$ to the second order in the
coupling constant $g$, $\la{\bf Q}\ra^{[2]}_{\rm ss}\! =\!{\bf
G}_0{\bf VG}_0{\bf VG}_0{\bf j}$, where ${\bf G}_0\equiv -{\bf
A}^{-1}$. Finally, the correlation functions relevant for the
evaluation of the total ladder and crossed contributions [see
Eqs.~(\ref{Cterm}), (\ref{Lterm})] read \beml \beq
\la\sigma^{1}_{21}\sigma^2_{12}\ra^{[2]}_{\rm ss}\!\!\!&=
&\!\!\!\la[{\bf Q}]_{152}\ra_{\rm ss}^{[2]},\\
\la\sigma^{1}_{12}\sigma^2_{21}\ra^{[2]}_{\rm ss}\!\!\!&
=&\!\!\!\la[{\bf Q}]_{137}\ra_{\rm ss}^{[2]},\\
2\la\sigma^{1}_{22}\ra^{[2]}_{\rm ss}\!\!\!& =&\!\!\!\la[{\bf
Q}]_{16}\ra_{\rm ss}^{[2]}\!+\!\la[{\bf
Q}]_{32}\ra_{\rm ss}^{[2]}\!+\!\la[{\bf Q}]_{48}\ra_{\rm ss}^{[2]},\\
2\la\sigma^{2}_{22}\ra^{[2]}_{\rm ss}\!\!\!& =&\!\!\!\la[{\bf
Q}]_{1}\ra_{\rm ss}^{[2]}\!+\! \la[{\bf Q}]_{2}\ra_{\rm
ss}^{[2]}\!+\!\la[{\bf Q}]_{3}\ra_{\rm ss}^{[2]}. \eq
\label{rel_corr} \eml Concerning the elastic ladder and crossed
terms [see Eq.~(\ref{el_part})], the relevant dipole moment
expectation values are given by \beml \beq
\la\sigma^{1}_{12}\ra^{[1]}_{\rm ss}\!&=&\!\la[{\bf
Q}]_{128}\ra^{[1]}_{\rm ss},\,\,\la\sigma^{1}_{21}\ra^{[1]}_{\rm
ss}\!=
\!\la[{\bf Q}]_{144}\ra^{[1]}_{\rm ss},\\
\la\sigma^{2}_{12}\ra^{[1]}_{\rm ss}\!&= &\!\la[{\bf
Q}]_{8}\ra^{[1]}_{\rm
ss},\,\,\,\,\,\,\la\sigma^{2}_{21}\ra^{[1]}_{\rm ss}\!=\!\la[{\bf
Q}]_{9}\ra^{[1]}_{\rm ss}, \eq \eml where $\la {\bf Q}\ra_{\rm
ss}^{[1]}={\bf G}_0{\bf VG}_0{\bf j}$.

Numerical results for the elastic and inelastic intensities
for different detunings can be found in the main text. In the next subsection,
we will present analytical results for the case of exact resonance.

\subsection{Analytical results for $\delta=0$}
Evaluating correlation functions from Eq.~(\ref{rel_corr}), we
arrive at following results \cite{shatokhin06}
\beq
2\re\{\la\sigma_{21}^1\sigma_{12}^2\ra_{\rm
ss}^{[2]}e^{i{\bf k}\cdot{\bf
r}_{12}}\}&=&|g|^2|\tD_{+1,+1}|^2\frac{R_1(s)}{(4+s)P(s)}\n\\
&&\times\cos\{({\bf k}+{\bf k}_L)\cdot{\bf r}_{12}\}\label{cohs},\\
\la\sigma_{22}^1\ra_{\rm ss}^{[2]}+\la\sigma_{22}^2\ra_{\rm
ss}^{[2]}&=&|g|^2|\tD_{+1,+1}|^2\frac{R_2(s)}{P(s)}\, . \label{pops}
\eq
$R_1(s)$, $R_2(s)$, and $P(s)$ are polynomial expressions in the
on-resonance saturation parameter $s=\Omega^2/2\gamma^2$,
\beml
\beq
R_1(s)&=&\frac{2}{9}\left(6912s+3168s^2\rt.\n\\
&&\lt.+264s^3+20s^4+s^5\rt),\\
R_2(s)&=&\frac{1}{3}\lt(1152s+528s^2+132s^3+7s^4\rt),\\
P(s)&=&(1+s)^2(12+s)(32+20s+s^2),
\eq
\label{RRP}
\eml
and $\tD_{+1,+1}=\unite_{+1}\cdot\tD\cdot\unite_{+1}$.

The configuration average of (\ref{cohs}) and (\ref{pops}) leads to
the final result \beq C^{\rm tot}_2(\theta)&\simeq&
\frac{|\bar{g}|^2R_1(s)}{(4+s)P(s)}\Bigl(\frac{2}{15}-
\frac{(k\,\ell\,\theta)^2}{35}\Bigr),
\label{CTheta}\\
L^{\rm tot}_2&=&\frac{2|\bar{g}|^2R_2(s)}{15P(s)},\label{LHparH} \eq
with $\bar{g} = g|_{r_{\alpha\beta}=\ell}$. The scattering angle
$\theta=2\arcsin\{|{\bf k}+{\bf k}_L|/2k_L\}\ll 1$ with respect to
the backscattering direction was assumed to be sufficiently small
herein.

The enhancement factor $\alpha(s)$, Eq.~(\ref{efactor}), deduced from
Eqs.~(\ref{CTheta}) and (\ref{LHparH})
reads
\be
\alpha(s)=1+\frac{R_1(s)}{(4+s)R_2(s)}\, ,
\label{enh_res}
\e
and $\alpha(0)=2.0$ in the weak field limit, as
it must be in the elastic scattering regime.
For small $s$, enhancement linearly decreases as $2-s/4$, in full
agreement with the diagrammatic theoretical result \cite{wellens04}
and in qualitative agreement with the result of Sr experiment
\cite{chaneliere03}. When $s$ increases further, $\alpha$
monotonically drops to an asymptotic value
$\lim_{s\to\infty}\alpha(s)=\alpha_{\infty}=23/21$
\cite{shatokhin06} which is strictly larger than unity, implying a
residual constructive interference in the deep saturation regime.

We will next show that this interference is due to inelastic photons
only. Indeed, we obtained the following result for the elastic
ladder and crossed terms \be L^{\rm el}_2=C^{\rm
el}_2=\frac{2|\bar{g}|^2}{15}\frac{s}{(1+s)^4}. \label{LCel} \e As
seen from Eq.~(\ref{LCel}) the elastic component shows perfect
contrast for all $s$. In particular, it is this component that
results in enhancement $\alpha=2$ for very small $s\to 0$. However,
in the deep saturation regime, this component decreases as $s^{-3}$,
while the counterparts of the total intensity, Eqs.~(\ref{CTheta}),
(\ref{LHparH}), as $s^{-1}$. Herefrom follows our conclusion about
the origin of the residual enhancement in the deep saturation
regime. Explicitly, the inelastic crossed and ladder terms obtained
by elementary substraction of Eq.~(\ref{LCel}) from
Eqs.~(\ref{CTheta}) and (\ref{LHparH}) read \beq C^{\rm
inel}_2&=&\frac{2|\bar{g}|^2}{15}\frac{20736s^2+\ldots+2s^7}{9(1+s)^2(4+s)P(s)},
\label{Cinel}\\
L^{\rm
inel}_2&=&\frac{2|\bar{g}|^2}{15}\frac{2016s^2+\ldots+7s^6}{3(1+s)^2P(s)},\label{Linel}
\eq where we have retained only the lowest and highest order terms
in the numerators, because these terms will be used to verify the
expressions for the spectra. Explicit values of other coefficients
are not important. Using (\ref{Cinel}) and (\ref{Linel}), it is easy
to verify that $\lim_{s\to\infty}C^{\rm inel}_2/L^{\rm
inel}_2=2/21=\alpha_\infty-1$.

\section{Derivation of CBS spectrum from Eq.~(\ref{meq})}
\label{appA}
\subsection{Elastic and inelastic spectra}
By virtue of the
quantum regression theorem \cite{scully97}, the temporal correlation functions
of a Markov process obey the same equation of motion as the
expectation values of operators, that is, Eq.~(\ref{meq}), but with
different initial conditions and a different free term. The latter can
be straightforwardly determined provided that the stationary solution to Eq.~(\ref{meq})
is known.

The four correlation functions appearing in the definition of the
spectrum [see Eqs.~(\ref{G_tau}), (\ref{sp_fixed})] can be extracted
from the following correlation functions: \be {\bf
s}_\alpha(t)\equiv\la\sigma_{21}^\alpha{\bf Q}(t)\ra_{\rm ss}\quad
(\alpha=1,2), \label{def_s_alpha} \e where ${\bf s}_\alpha(t)$ is a
vector containing, like $\la{\bf Q}(t)\ra$, 255 elements. The vector
${\bf s}_\alpha(t)$ satisfies the equation of motion \be \dot{\bf
s}_\alpha=({\bf A}+{\bf V}){\bf
s}_\alpha+\la\sigma_{21}^\alpha\ra_{\rm ss}{\bf j}, \label{eqcorrf}
\e which explicitly accounts for the modification of the free term.
From the definition (\ref{def_s_alpha}), it follows that the vectors
${\bf s}_1$ and ${\bf s}_2$ are obtained from $\la{\bf
q}^{\prime}_1\otimes{\bf q}_2\ra_{\rm ss}$ and $\la{\bf
q}_1\otimes{\bf q}^{\prime}_2\ra_{\rm ss}$, respectively, where \beq
{\bf
q}^{\prime}&=&(\sigma_{21}/2,-\sigma_{21}/2,\sigma_{21}/2,-\sigma_{21}/2,
\sigma_{24},
0,\sigma_{23},0,\n\\
&&\sigma_{22},0,0,0,0,0,0,0)^T, \label{Qprime} \eq
by using the same mapping rule as described previously in Sec.~\ref{subsub:der_proc}.
Finally, to find the initial conditions for these vectors, one just needs to make
certain transpositions among the elements
of the vector $\la{\bf Q}\ra_{\rm ss}$.

The Laplace transform solution of Eq.~(\ref{eqcorrf}) reads: \be
\tilde{\bf s}_\alpha(z)=\frac{1}{z\openone_{255}-({\bf A}+{\bf
V})}\Bigl[{\bf s}_\alpha(0)+ \frac{\bf
j}{z}\la\sigma_{21}^\alpha\ra_{\rm ss}\Bigr], \label{sol_Laplace}
 \e
where $\tilde{\bf s}_\alpha(z)$ is a Laplace image of ${\bf
s}_\alpha(t)$. Using this solution and definitions \be
\la\sigma_{12}^1\ra_{\rm ss}=2\la [{\bf q}_1]_8\otimes [{\bf
q}_2]_0\ra, \,\, \la\sigma_{12}^2\ra_{\rm ss}= 2\la [{\bf
q}_1]_0\otimes [{\bf q}_2]_8\ra, \e we arrive at the following
expression for the Laplace image of the correlation function \beq
\tilde{G}^{(1)}_{\rm ss}(z)&=&2([\tilde{\bf
s}_1(z)]_{128}+[\tilde{\bf s}_2(z)]_{8}\n\\
&&+ [\tilde{\bf
s}_1(z)]_8e^{i{\bf k}\cdot{\bf r}_{12}}+[\tilde{\bf
s}_2(z)]_{128}e^{-i{\bf k}\cdot{\bf r}_{12}}), \label{G11z} \eq

The next step will be to extract the lowest-order in the coupling
$g$, nonvanishing contribution from (\ref{sol_Laplace}) and insert
it into Eq.~(\ref{G11z}). This contribution is on the order $g^2$
and corresponds to double scattering in which the atoms exchange a
photon. Explicitly, \beq \tilde{\bf s}_\alpha^{[2]}(z)&=&{\bf
G}_0(z){\bf V} {\bf G}_0(z){\bf V}{\bf G}_0(z)\Bigl[{\bf
s}_\alpha^{[0]}(0)+\frac{\bf j}{z}\la\sigma_{21}^\alpha\ra^{[0]}_{\rm ss}\Bigr]\n\\
&&+{\bf G}_0(z){\bf V}{\bf G}_0(z)\Bigl[{\bf
s}_\alpha^{[1]}(0)+\frac{\bf
j}{z}\la\sigma_{21}^\alpha\ra^{[1]}_{\rm ss}\Bigr]\n\\
&&+{\bf G}_0(z)\Bigl[{\bf s}_\alpha^{[2]}(0)+\frac{\bf j}{z}
\la\sigma_{21}^\alpha\ra^{[2]}_{\rm ss}\Bigr], \label{whole} \eq
where ${\bf G}_0(z)=(z\openone_{255}-{\bf A})^{-1}$, and the
superscripts $[0]$, $[1]$, and $[2]$ indicate terms on the order
$g^0$, $g^1$, and $g^2$, respectively.

Expression (\ref{whole}) can be simplified since (\ref{G11z}) gives
a spectrum in the $h\parallel h$ channel. Therefore, the stationary
expectation values of operators related to the laser-nondriven
transition vanish in $g^0$. So, the first, second, and last terms
must be dropped from (\ref{whole}). Concerning the latter term, it
does not contribute to (\ref{G11z}) since $[{\bf G}_0(z){\bf j}]_8$
and $[{\bf G}_0(z){\bf j}]_{128}$ give
$\la\sigma_{12}^2\ra^{[0]}_{\rm ss}$ and
$\la\sigma_{12}^1\ra^{[0]}_{\rm ss}$, respectively, and thus vanish
for the reason already indicated.

It is useful to split the nonvanishing part of (\ref{whole}) into
two counterparts from which the elastic and inelastic spectra are
extracted: \be \tilde{\bf s}_\alpha^{[2]}(z)=\tilde{\bf
s}_{\alpha;\rm el}^{[2]}(z)+ \tilde{\bf s}_{\alpha;\rm
inel}^{[2]}(z), \e where the two vectors \beml \beq \tilde{\bf
s}_{\alpha;\rm el}^{[2]}(z)&=&\frac{1}{z}{\bf G}_0{\bf V}{\bf
G}_0{\bf
j}\la\sigma_{21}^\alpha\ra^{[1]}_{\rm ss},\label{elastic}\\
\tilde{\bf s}_{\alpha;\rm inel}^{[2]}(z)&=&{\bf G}_0(z){\bf V}
{\bf G}_0(z){\bf s}_\alpha^{[1]}(0)
+{\bf G}_0(z){\bf s}_\alpha^{[2]}(0)\n\\
&&+\frac{\la\sigma_{21}^\alpha\ra^{[1]}_{\rm ss}}{z}\lt[{\bf G}_0(z)
{\bf V}{\bf G}_0(z)-
{\bf G}_0{\bf V}{\bf G}_0\rt]{\bf j},\n\\
\label{inelastic}
\eq
\eml
lead to the elastic and inelastic spectra, respectively.

Inserting (\ref{elastic}) into (\ref{G11z}) we obtain
\beq
[\tilde{G}^{(1)}_{\rm ss}(z)]_{\rm el}&=&\frac{1}{z}\lt(\la\sigma_{21}^1\ra^{[1]}_{\rm ss}
\la\sigma_{12}^1\ra_{\rm ss}^{[1]}+\la\sigma_{21}^2\ra^{[1]}_{\rm ss}
\la\sigma_{12}^2\ra_{\rm ss}^{[1]}\rt.\n\\
&&\lt.+2\re\{\la\sigma_{21}^1\ra^{[1]}_{\rm ss}
\la\sigma_{12}^2\ra_{\rm ss}^{[1]}e^{i{\bf k}\cdot{\bf
r}_{12}}\}\rt), \label{s_el_z} \eq with the right hand side
expression in the round brackets being nothing but the stationary
elastic intensity $I_2^{\rm el}$, Eq.~(\ref{el_part}). Finally,
putting $[\tilde{G}^{(1)}_{\rm ss}(z)]_{\rm el}$ to (\ref{sp_fixed})
we arrive at \be \tilde{I}_2^{\rm el}(\nu)=I_2^{\rm el}\delta(\nu),
\label{I2nu_el} \e where we have used the formula \be
\lim_{\Gamma\to
0}\frac{1}{\Gamma-i\nu}=\pi\delta(\nu)+iP\frac{1}{\nu}, \e with $P$
denoting the principal value of an integral.

One can check that the right hand side of Eq.~(\ref{inelastic}) does
not have a pole at $z=0$, which means that the respective expression
describes the inelastic spectrum.

\subsection{Analytical results for inelastic spectrum at $\delta=0$}
\subsubsection{Weak field ($\Omega\ll\gamma$)}
At small Rabi frequencies, the analytical formulas are obtained
after taking into account the lowest-order inelastic process -- the
two-photon scattering, -- and neglecting the inelastic processes of
higher orders. The two-photon processes are proportional to the
square of the intensity, that is, to $\Omega^4$. The ladder and
crossed terms read (we omit the common prefactor $2|\bar{g}|^2/15$):
\beq \tilde{L}^{\rm
inel}_2(\nu)&\simeq&\frac{1}{\pi}\lt(\frac{\Omega}{\gamma}\rt)^4
\frac{\gamma^3(2\gamma^2+\nu^2)}{2(\gamma^2+\nu
^2)^3},\label{LCina0}\\
\tilde{C}^{\rm
inel}_2(\nu)&\simeq&\frac{1}{\pi}\lt(\frac{\Omega}{\gamma}\rt)^4
\frac{\gamma^5}{(\gamma^2+\nu^2)^3}. \label{LCina} \eq
It is easy to check that the expressions (\ref{LCina0}), (\ref{LCina}) are consistent with the behavior of the 
enhancement factor in the two-photon scattering regime. Integrating
$\tilde{L}^{\rm
inel}_2(\nu)$, $\tilde{C}^{\rm
inel}_2(\nu)$ over all frequencies, we obtain the
following inelastic ladder and crossed terms for small $\Omega$: \beq
L^{\rm inel}_2&=&\int^{\infty}_{-\infty} d\nu\tilde{L}^{\rm
inel}_2(\nu)=\frac{7}{16}\lt(\frac{\Omega}{\gamma}\rt)^4,\label{LandCin0}\\
C^{\rm inel}_2&=&\int^{\infty}_{-\infty} d\nu\tilde{C}^{\rm
inel}_2(\nu)=\frac{3}{8}\lt(\frac{\Omega}{\gamma}\rt)^4.
\label{LandCin}
\eq Combining Eqs.~(\ref{LandCin0}), (\ref{LandCin}) with the small-$s$
expression for the elastic ladder and crossed terms $L^{\rm
el}_2=C^{\rm el}_2=s$, and rewriting Eq.~(\ref{LandCin}) in terms of
$s$, we recover the expected linear decrease \be
\alpha=1+\frac{s+3s^2/2}{s+7s^2/4}\simeq 2-\frac{s}{4}.
\e
\subsubsection{Strong field ($\Omega\gg\gamma$)}
In the opposite limit of a strong field,
the CBS intensity is inversely proportional to
the laser field intensity. We will now present
the analytical expressions for the ladder and crossed spectra
derived in the leading order $\sim (\gamma/\Omega)^2$.

In this case, explicit expressions for CBS spectra can be
represented by using a function of two real variables $x_1$ and
$x_2$: \be \pounds(x_1,x_2)=\frac{1}{\pi}\frac{x_1}{x_1^2+x_2^2}.
\label{pounds} \e Let us mention the properties of
$\pounds(x_1,x_2)$ that are important to us: (i) if $x_1={\rm
Const}$, then the function (\ref{pounds}) represents a Lorentzian
with full width at half maximum (referred to as {\it width} in the
main text) $2x_1$ and resonance at $x_2=0$; (ii) if $x_2={\rm
Const}$, then (\ref{pounds}) describes a resonance of a dispersive
type at $x_1=0$, with the width $2x_2$.

With the help of the function (\ref{pounds}), the ladder and crossed
spectra are given by
\beq \tilde{L}^{\rm
inel}_2(\nu)\!\!&\simeq&\!\!\lt(\frac{\gamma}{\Omega}\rt)^2\Bigl(\frac{1}{2}\pounds(\gamma,\nu)+
\frac{1}{4}\pounds(3\gamma,\nu)\Bigr.\n\\
&&\Bigl.+\frac{14}{9}[\pounds(3\gamma/2,\nu-\Omega/2)+\pounds(3\gamma/2,\nu+\Omega/2)]\Bigr.\n\\
&&\Bigl.+\frac{1}{9}[\pounds(3\gamma/2,\nu-\Omega)+\pounds(3\gamma/2,\nu+\Omega)]\Bigr.\n\\
&&\Bigl.+\frac{5}{18}[\pounds(5\gamma/2,\nu-\Omega)+\pounds(5\gamma/2,\nu+\Omega)]\Bigr.\n\\
&&\Bigl.+\frac{1}{72}[\pounds(3\gamma,\nu-2\Omega)+\pounds(3\gamma,\nu+2\Omega)]\Bigr),
\label{lad:asymp}\\ \tilde{C}^{\rm
inel}_2(\nu)\!\!&\simeq&\!\!\lt(\frac{\gamma}{\Omega}\rt)^2\Bigl(\frac{1}{2}\pounds(2\gamma,\nu)+
\frac{1}{4}\pounds(3\gamma,\nu)\Bigr.\n\\
&&\Bigl.-\frac{1}{6}[\pounds(5\gamma/2,\nu-\Omega)+
\pounds(5\gamma/2,\nu+\Omega)]\Bigr.\n\\
&&\Bigl.+\frac{1}{72}[\pounds(3\gamma,\nu-2\Omega)+\pounds(3\gamma,\nu+2\Omega)]\Bigr)\n\\
&&+\lt(\frac{\gamma}{\Omega}\rt)^3\frac{208}{45}\n\\
&&\times[\pounds(\nu+\Omega/2,3\gamma/2)
-\pounds(\nu-\Omega/2,3\gamma/2)],\n\\
\label{cro:asymp} \eq where the
two terms of order $(\gamma/\Omega)^3$ are retained because they
define dispersive resonances of $\tilde{C}^{\rm inel}_2(\nu)$ at
$\nu=\pm\Omega/2$.
By performing the elementary integrations of Eqs.~(\ref{lad:asymp}) and
(\ref{cro:asymp}) we arrive at the inelastic ladder and crossed
terms \be L^{\rm inel}_2\simeq
\frac{14}{3}\lt(\frac{\gamma}{\Omega}\rt)^2,\quad C^{\rm
inel}_2\simeq \frac{4}{9}\lt(\frac{\gamma}{\Omega}\rt)^2,
\label{LCtot} \e which are consistent with Eqs.~(\ref{Cinel}), (\ref{Linel}) and, hence, with 
$\alpha=\alpha_\infty=23/21$.


\begin{thebibliography}{99}

\bibitem{kupriyanov06}
D.~V.~Kupriyanov, I.~M.~Sokolov, C.~I.~Sukenik, and  M.~D.~Havey,
Laser Phys. Lett. {\bf 3}, 223 (2006).

\bibitem{labeyrie99}
G.~Labeyrie, F.~de Tomasi, J.-C. Bernard, C.~A.~M\"uller, C.~Miniatura
and R.~Kaiser, Phys. Rev. Lett. {\bf 83}, 5266 (1999).

\bibitem{kulatunga03}
P.~Kulatunga, C.~I.~Sukenik, S.~Balik, M.~D.~Havey,
D.~V.~Kupriyanov, and I.~M.~Sokolov, Phys.\ Rev.
A {\bf 68}, 033816 (2003).

\bibitem{meso94}
E.~Akkermans, G.~Montambaux, J.-L.~Pichard, and J.~Zinn-Justin (Eds.) {\it Mesoscopic Quantum
Physics} (Elsevier, Amsterdam, 1994).

\bibitem{lukin03}
M.~D.~Lukin, Rev.~Mod.~Phys. {\bf 75}, 457 (2003).


\bibitem{datsuk06}
V.~M.~Datsyuk, I.~M.~Sokolov, D.~V.~Kupriyanov, and M.~D.~Havey,
Phys.~Rev.~A {\bf 74}, 043812 (2006).

\bibitem{skipetrov06}
S.~E.~Skipetrov, Phys.~Rev. A {\bf 75}, 053808 (2007).

\bibitem{chaneliere03}
T.~Chaneli\`ere, D.~Wilkowski, Y.~Bidel, R.~Kaiser,
and C.~Miniatura, Phys.\ Rev. E {\bf 70}, 036602 (2004).

\bibitem{balik05}
S.~Balik, P.~Kulatunga, C.~I.~Sukenik,~M.~D.~Havey, D.~V.~Kupriyanov, and I.~M.~Sokolov, J. Mod. Opt.
{\bf 52}, 2269 (2005).

\bibitem{jonckheere00}
T.~Jonckheere, C.~A.~M\"uller, R.~Kaiser, C.~Miniatura, and D.~Delande,
Phys.~Rev.~Lett. {\bf 85}, 4269 (2000).

\bibitem{mueller01}
C.~A.~M\"uller, T.~Jonckheere, C.~Miniatura, and D.~Delande,
Phys.\ Rev. A {\bf 64}, 053804 (2001).

\bibitem{mueller02}
C.~A.~M\"uller and C.~Miniatura, J.~Phys. A {\bf 35}, 10163 (2002).

\bibitem{kupriyanov03}
D.~V.~Kupriyanov, I.~M.~Sokolov, P.~Kulatunga, C.~I.~Sukenik, and  M.~D.~Havey,
Phys.\ Rev. A {\bf 67}, 013814 (2003).


\bibitem{labeyrie03}
G.~Labeyrie, D.~Delande, C.~A.~M\"uller, C.~Miniatura, and
R.~Kaiser, Europhys. Lett. {\bf 61}, 327 (2003).


\bibitem{bidel02}
Y.~Bidel, B.~Klappauf, J.~C.~Bernard, D.~Delande, G.~Labeyrie, C.~Miniatura,
D.~Wilkowski, and R.~Kaiser, Phys.~Rev.~Lett. {\bf 88}, 203902 (2002).


\bibitem{wellens04}
T.~Wellens, B.~Gr\'emaud, D.~Delande, and C.~Miniatura, Phys. Rev. A {\bf 70},
023817 (2004).

\bibitem{wellens05}
T.~Wellens, B.~Gr\'emaud, D.~Delande, and C.~Miniatura, Phys. Rev. E
{\bf 71}, 055603(R) (2005).

\bibitem{wellens06}
T.~Wellens, B.~Gr\'emaud, D.~Delande, and C.~Miniatura, Phys. Rev. A
{\bf 73}, 013802 (2006).

\bibitem{shatokhin05}
V. Shatokhin, C.~A.~M\"uller, and A. Buchleitner, \prl {\bf 94},
043603 (2005).

\bibitem{shatokhin06}
V. Shatokhin, C.~A.~M\"uller, and A. Buchleitner, \pra {\bf 73},
063813 (2006).

\bibitem{gremaud06}
B.~Gr\'emaud, T.~Wellens, D.~Delande, C.~Miniatura, Phys.~Rev.~A
{\bf 74}, 033808 (2006).

\bibitem{shatokhin07}
V. Shatokhin, Opt. Spectrosc. {\bf 103}, 300 (2007); preprint
arXiv:quant-ph/0608094 (2006).


\bibitem{glauber65}
R.~J.~Glauber, in {\it Quantum Optics and Electronics}, edited by
C.~DeWitt, A.~Blandin, and C. Cohen-Tannoudji (Gordon and Breach,
London, 1965).

\bibitem{scully97}
M.~O.~Scully and M.~S.~Zubairy, {\it Quantum Optics}, (Cambridge
University Press, Cambridge, U.~K., 1997).

\bibitem{lehmberg70}
R.~H.~Lehmberg, Phys.\ Rev. A {\bf 2}, 883 (1970).

\bibitem{mollow69}
B.~R.~Mollow, Phys.\ Rev. {\bf 188}, 1969 (1969).

\bibitem{shirley65}
J.~H.~Shirley, Phys. Rev. {\bf 138}, B979 (1965).

\bibitem{cohen_tannoudji}
C.~Cohen-Tannoudji, J.~Dupont-Roc, G.~Grynberg, {\it Atom-Photon
Interactions} (Wiley, New York, 1992).

\bibitem{Boyd}
R.~W.~Boyd, {\it Nonlinear optics} (Academic, San Diego, 1992).

\bibitem{townes55}
S.~H.~Autler and C.~H.~Townes, Phys.~Rev. {\bf 100}, 703 (1955).

\bibitem{mollow72b}
B.~R.~Mollow, Phys.~Rev.~A {\bf 5}, 1522 (1972).

\bibitem{shatokhin01}
V.N. Shatokhin and S.Ya. Kilin, Phys. Rev. A {\bf 63}, 023803 (2001).

\bibitem{kupriyanov04}
{D.V. Kupriyanov, I.M. Sokolov, and M.D. Havey} Opt. Commun.
\textbf{243}, 165 (2004).

\end{thebibliography}
\end{document}